\documentclass[aps,prd,twocolumn,superscriptaddress,nofootinbib]{revtex4-1}

\usepackage{amsmath, amssymb}
\usepackage{slashed}

\newcommand{\Sf}{S_F}
\newcommand{\dL}{\displaystyle\int\frac{d^d k}{(2\pi)^d}}

\begin{document}

\title{Gauge-Invariant Fermion}
\author{Kang-Sin Choi}
\email{kangsin@ewha.ac.kr}
\affiliation{Scranton Honors Program, Ewha Womans University, Seoul 03760, Korea}
\affiliation{Institute of Mathematical Sciences, Ewha Womans University, Seoul 03760, Korea}

\begin{abstract}
We show that in quantum electrodynamics the Green's functions of the Dirac-dressed fermion coincide, for any value of the gauge parameter in
the $R_\xi$ gauge and to all orders in perturbation theory, with those of the undressed fermion in the Landau gauge.
The dressed fermion couples only to the transverse photon, and the gauge parameter is confined to a decoupled sector.
The dressed vertex then obeys a homogeneous Ward identity, and the dressed field is renormalized multiplicatively with no new counterterm.
A partially dressed family realizes the Landau--Khalatnikov--Fradkin relation exactly, and dressings with other kernels reproduce other
gauges. Every dressing yields the same pole mass and, with the differences carried by the vertex, the same mass function.
\end{abstract}

\maketitle

\section{Introduction}

A gauge-invariant charged field was first constructed by
Dirac~\cite{Dirac:1955}, who attached a ``photon cloud'' to the bare
electron field,
\begin{equation}
  \psi_D(x) = e^{ie\chi(x)}\psi(x).
  \label{eq:Dirac_field}
\end{equation}
Dirac's original choice $\chi = \partial_i A_i/\nabla^2$ creates the
electron together with its Coulomb field. In this paper we use the
Lorentz-covariant kernel
\begin{equation}
  \chi(x) = \frac{\partial_\mu A^\mu}{\partial^2}(x),
  \label{chi}
\end{equation}
which is the dressing associated with the Lorenz condition~\cite{Lavelle:McMullan:1996,Lavelle:McMullan:1995,%
Berghofer:Francois:2024}. Following common usage, we refer to
Eqs.~\eqref{eq:Dirac_field} and~\eqref{chi} as the Dirac dressing.
Under a gauge transformation
$A_\mu\to A_\mu+\partial_\mu\Lambda$, $\psi\to e^{-ie\Lambda}\psi$, the
dressing transforms as $\chi\to\chi+\Lambda$, so $\psi_D$ is gauge
invariant. We call $\psi_D$ the dressed fermion.

The inverse d'Alembertian makes $\psi_D$ nonlocal in spacetime.
This raises the questions whether the dressed field can be treated with
the standard tools of perturbative quantum field theory, whether it
requires a separate composite-operator
renormalization~\cite{Frohlich:1981yi} and how it
depends on the choice of
dressing~\cite{Lavelle:McMullan:1995,Bagan:Lavelle:McMullan:2000}. See
also Ref.~\cite{Grassi:Porrati:2024} for a recent operator-algebraic
approach to charged states and Ref.~\cite{Berghofer:Francois:2024} for
the relation between dressings and gauge fixings.

It is known that a gauge fixing can be used to construct a dressing,
and that the dressed field reduces to the undressed one in the
corresponding gauge~\cite{Lavelle:McMullan:1996,Lavelle:McMullan:1995}.
For the Dirac dressing~\eqref{chi}, this gauge is the Landau gauge.
One-loop propagators of dressed electrons were computed in
Refs.~\cite{Bagan:Lavelle:McMullan:1996,Bagan:1996su,%
Bagan:Lavelle:McMullan:2000}, where it was also shown that dressings
adapted to the velocity of the charge admit an infrared-finite
mass-shell renormalization. Propagators in different covariant gauges
are related by the Landau--Khalatnikov--Fradkin (LKF)
transformation~\cite{Landau:Khalatnikov:1955,Fradkin:1955,%
Johnson:Zumino:1959,Zumino:1959wt}. The propagator of the dressed
fermion equals the Landau-gauge propagator, and the LKF transformation
follows from a field-dependent gauge transformation that rescales the
gauge parameter~\cite{DeMeerleer:2018}. In non-Abelian theories, the Green's
functions of a suitably dressed quark field are independent of the
gauge parameter~\cite{DeMeerleer:2018,DeMeerleer:2020,%
Capri:2019drm}.

In this paper we give an explicit all-orders proof of these relations
for the Dirac dressing in QED, covering all Green's functions of the
dressed fermion and the transverse photon. Our main result is the
following statement.

\medskip
\noindent\textit{Proposition.} In perturbative QED, for \emph{any value
of $\xi$} in the $R_\xi$ gauge, the Green's functions of $\psi_D$,
$\overline{\psi}_D$ and the transverse photon $A^T_\mu$ coincide with the
Green's functions of $\psi$, $\overline \psi$ and $A_\mu$ in the Landau gauge,
$\xi=0$.
\medskip

The proof in Sec.~\ref{sec:allorders} is an exact change of variables
in the regularized path integral of local QED. It does not take the
nonlocal transformed Lagrangian as a starting point. The consequences
are derived in the subsequent sections. The dressed photon vertex obeys
a homogeneous Ward identity, and the dressed field is renormalized
multiplicatively by the Landau-gauge wave-function renormalization,
with no new counterterm and no operator mixing (Sec.~\ref{sec:WI}). A
one-parameter family of partially dressed fields realizes the LKF
transformation exactly (Sec.~\ref{sec:LKF}). Dressings with other
kernels reproduce the Green's functions of other gauges, and a
gauge-invariant Wilson-line bilocal is a different object whose contour
dependence is physical (Sec.~\ref{sec:other}). All dressings share the
pole mass. Off shell, the mass function of the dressed field has the
Landau form. Assigning its difference from the Feynman form to the
vertex, as in the pinch technique, yields the gauge-invariant mass
function of Refs.~\cite{Choi:2026yxc,Choi:2026edn}
(Sec.~\ref{sec:mass}).

We restrict the analysis to QED. The dressing remains nonlocal, and we
limit our discussion of $\psi_D$ to its Green's functions.

\section{Transverse coupling of the dressed field}
\label{sec:transverse}

We define the inverse d'Alembertian in Euclidean space, where it is
unambiguous for fields that vanish at infinity, as in the Euclidean
construction of the gauge-invariant transverse field in
Ref.~\cite{DeMeerleer:2018}. After continuation to Minkowski
space~\cite{Wick:1954,Osterwalder:Schrader:1973,%
Osterwalder:Schrader:1975} this is the Feynman prescription,
$1/\partial^2\to -1/(k^2+i\epsilon)$ in momentum space, the same
prescription that defines the projector in the Landau-gauge photon
propagator. A retarded or advanced Green's function would define a
different operator, to which the results below need not apply.

Substituting $\psi = e^{-ie\chi}\psi_D$ into the QED Lagrangian and
using $i\partial_\mu(e^{-ie\chi}\psi_D) = e^{-ie\chi}(i\partial_\mu +
e\partial_\mu\chi)\psi_D$, we obtain
\begin{equation}
  \overline \psi(i\slashed\partial - e\slashed A - m)\psi
  = \overline{\psi}_D(i\slashed\partial - e\gamma^\mu A^T_\mu - m)\psi_D,
  \label{eq:dressed_Lagrangian}
\end{equation}
where
\begin{equation}
  A^T_\mu = P^T_{\mu\nu}A^\nu, \qquad
  P^T_{\mu\nu} = g_{\mu\nu} - \frac{\partial_\mu\partial_\nu}{\partial^2}.
  \label{eq:AT}
\end{equation}
The dressed fermion couples only to the transverse photon. The
projector $P^T_{\mu\nu}$, and with it the interaction in
Eq.~\eqref{eq:dressed_Lagrangian}, is nonlocal in spacetime, so
Eq.~\eqref{eq:dressed_Lagrangian} is not a conventional local
Lagrangian density. We use it only as an algebraic identity and base
the all-orders argument of Sec.~\ref{sec:allorders} on local QED.

Reading off the interaction in Eq.~\eqref{eq:dressed_Lagrangian} in momentum space, where $\partial_\mu\to -i k_\mu$ and
$\partial^2\to -k^2$, the tree-level dressed vertex coupling to $A_\mu$ with photon momentum $k$ is $-ie\Gamma_{D,0}^\mu$ with \cite{Lavelle:McMullan:1995,Bagan:Lavelle:McMullan:2000}
\begin{equation}
  \Gamma_{D,0}^\mu \equiv \gamma_\nu P^{T\nu\mu}(k) \equiv \gamma^\mu - \frac{\slashed k k^\mu}{k^2}.
  \label{eq:dressed_vertex}
\end{equation}
The subscript $0$ distinguishes this tree-level vertex from the full one-particle irreducible (1PI) vertex $\Gamma_D^\mu$ that we deal with later. 
The coupling $e$ is unchanged; the dressing modifies the Lorentz structure, not the coupling strength. The tree-level vertex is transverse,
\begin{equation}
  \Gamma_{D,0}^\mu k_\mu = \slashed k - \frac{\slashed k\, k^2}{k^2} = 0.
  \label{transv}
\end{equation}
In Sec.~\ref{sec:WI} we show that this is the tree-level form of an
exact Ward identity.

\section{One-loop self-energy}
\label{sec:oneloop}

With the tree-level vertex~\eqref{eq:dressed_vertex}, the one-loop
self-energy of the dressed fermion is
\begin{equation}
  -i\Sigma^{(\psi_D)}(q)
  = (-ie)^2\dL\,
    \Gamma_{D,0}^\mu\Sf(q{-}k)\Gamma_{D,0}^\nu D_{\mu\nu}(k),
  \label{eq:dressed_selfenergy}
\end{equation}
where $\Sf(k) = i/(\slashed{k}-m)$ and the $R_\xi$ photon propagator
splits into transverse and longitudinal parts,
\begin{equation}
\begin{split}
 D_{\mu\nu}(k) &= D^T_{\mu\nu}(k) + D^L_{\mu\nu}(k), \\
 D^T_{\mu\nu}(k) &\equiv -\frac{i}{k^2}\left(g_{\mu\nu}
   - \frac{k_\mu k_\nu}{k^2}\right), \quad
 D^L_{\mu\nu}(k) \equiv -i\xi\,\frac{k_\mu}{k^2}\frac{k_\nu}{k^2}.
\end{split}
\label{eq:photon_split}
\end{equation}
The longitudinal part factorizes into one factor $k_\mu/k^2$ for each
end of the photon line.\footnote{Each factor $1/k^2$ is the inverse
d'Alembertian of the dressing~\eqref{chi} under the replacement
$\partial^2\to -k^2$ of Sec.~\ref{sec:transverse}. Since
$k^\mu k^\nu D_{\mu\nu}(k) = -i\xi$ is momentum independent,
$D^L_{\mu\nu}$ is the propagator of $\partial_\mu\chi$, with one inverse
d'Alembertian from $\chi$ at each end of the line
(Sec.~\ref{sec:allorders}).} At each end the dressing absorbs the factor
$k_\mu/k^2$, $\Gamma_{D,0}^\mu k_\mu/k^2 = 0$ by Eq.~\eqref{transv}, so
\begin{equation}
  \Gamma_{D,0}^\mu\,\Sf(q{-}k)\,\Gamma_{D,0}^\nu\,D^L_{\mu\nu}(k) = 0.
  \label{eq:DP_vanishes}
\end{equation}
Only the transverse propagator contributes. Since
$D^T_{\mu\nu}k^\nu = 0$, the dressing correction $-\slashed k k^\mu/k^2$
in $\Gamma_{D,0}^\mu$ also drops out,
$\Gamma_{D,0}^\mu D^T_{\mu\nu} = \gamma^\mu D^T_{\mu\nu}$. At one loop, the
self-energy of the dressed field is therefore
\begin{equation}
  \hat\Sigma^{(\psi_D)}(q) = \Sigma_{\xi=0}(q),
  \label{eq:result}
\end{equation}
the conventional self-energy in the Landau gauge. Here and below a hat
denotes a self-energy that does not depend on the gauge parameter.
Equation~\eqref{eq:result} is the result of this explicit one-loop
calculation. Its extension to all orders follows separately from the
argument of Sec.~\ref{sec:allorders}.

The subscript $\xi=0$ labels the form of the result, not the gauge used
in the computation. The calculation is performed in an arbitrary
$R_\xi$ gauge, and the transversality of $\Gamma_{D,0}^\mu$ projects the
answer onto $\Sigma_{\xi=0}$ regardless of $\xi$. No gauge condition is
imposed, so this is not gauge fixing. The Landau form is an output of
the calculation, not an input. Following Ref.~\cite{Choi:2026edn} we
write the conventional one-loop self-energy as
\begin{equation}
  \Sigma_\xi(q) = \Sigma_{\xi=1}(q) + (1-\xi)(\slashed q - m)\,\Omega(q).
  \label{eq:Omega}
\end{equation}
Here $(1-\xi)(\slashed q - m)\,\Omega(q)$ is the contribution of the
term $i(1-\xi)k_\mu k_\nu/(k^2)^2$, by which the $R_\xi$ propagator
differs from the Feynman-gauge one. Applying the Ward--Takahashi
identity $\slashed k = (\slashed q - m) - (\slashed q - \slashed k - m)$
at both vertices shows that this contribution is proportional to
$\slashed q - m$, with the coefficient
\begin{equation*}
  \Omega(q) = ie^2\dL\,\frac{1}{(k^2)^2}
  \left[\frac{1}{\slashed q - \slashed k - m}\,(\slashed q - m)
  - 1\right].
\end{equation*}
Thus $\Omega(q)$ is a combination of the unit matrix and $\slashed q$
with coefficients depending on $q^2$, and it is ultraviolet divergent.
Its explicit form in terms of standard one-loop functions is given in
Ref.~\cite{Choi:2026edn}. Equation~\eqref{eq:Omega} is a one-loop
relation, since beyond one loop the self-energy contains higher powers
of $\xi$, and the same holds for the relations derived from it below.
The longitudinal contribution
$\Sigma_\xi - \Sigma_{\xi=0} = -\xi(\slashed q - m)\,\Omega(q)$, which
vanishes at the pole, is what the transversality of the tree-level
vertex eliminates.

\section{Equivalence to all orders}
\label{sec:allorders}

The result~\eqref{eq:result} extends to all orders and to all Green's
functions of the dressed fields. We start from local QED in the
$R_\xi$ gauge with dimensional regularization and couple sources to the
gauge-invariant fields $\psi_D$, $\overline{\psi}_D$ and $A^T_\mu$,
\begin{equation}
\begin{split}
  Z[\eta,\bar\eta,J] = \int &\mathcal{D}\psi\,\mathcal{D}\overline \psi\,
  \mathcal{D}A\,\exp\Big\{iS_\xi \\
  &+ i\!\int\! d^dx\,
  (\bar\eta\psi_D + \overline{\psi}_D\eta + J^\mu A^T_\mu)\Big\},
\end{split}
  \label{eq:Z}
\end{equation}
where
\begin{equation}
\begin{split}
  S_\xi = \int d^dx\Big[&\overline \psi(i\slashed\partial - e\slashed A - m)\psi \\
  &- \frac14 F_{\mu\nu}F^{\mu\nu}
  - \frac{1}{2\xi}(\partial_\mu A^\mu)^2\Big].
\end{split}
  \label{eq:S_xi}
\end{equation}
We change variables to
\begin{equation}
  A_\mu = A^T_\mu + \partial_\mu\chi, \quad
  \psi = e^{-ie\chi}\psi_D, \quad
  \overline \psi = \overline{\psi}_D\, e^{ie\chi},
  \label{eq:change}
\end{equation}
with $\partial^\mu A^T_\mu = 0$ and $\chi$ given by Eq.~\eqref{chi}.
The decomposition of the gauge field in Eq.~\eqref{eq:change} is an
identity for every gauge-field configuration that vanishes at infinity,
not a gauge condition. Under a gauge transformation, $A^T_\mu$ is
invariant and $\chi\to\chi+\Lambda$, so $A^T_\mu$ labels the gauge
orbit and $\chi$ the position along it. The path integral still runs
over all configurations, and the gauge is fixed only by the $R_\xi$
term, which in these variables weights $\chi$ alone. The Landau-gauge
configurations $\chi=0$ are reached only in the limit $\xi\to0$ of this
weight, which we use below to evaluate Green's functions already shown to
be independent of $\xi$. In the language of
Ref.~\cite{Berghofer:Francois:2024}, $A^T_\mu$ is a dressed field
rather than a gauge-fixed potential.

The change of the gauge field is linear, with a field-independent Jacobian. 
The fermionic Jacobians of $\psi$ and $\overline \psi$ cancel, since the transformation is a local vector phase rotation having no anomaly.
The field strength depends only on $A^T_\mu$, the gauge-fixing term depends only on $\partial_\mu A^\mu = \partial^2\chi$ and the fermion Lagrangian becomes Eq.~\eqref{eq:dressed_Lagrangian}. The action therefore separates,
\begin{equation}
  S_\xi = S_{\rm T}[\psi_D,\overline{\psi}_D,A^T]
  - \frac{1}{2\xi}\int d^dx\,(\partial^2\chi)^2,
  \label{eq:split}
\end{equation}
with
\begin{equation}
\begin{split}
  S_{\rm T} = \int d^dx\Big[&\overline{\psi}_D(i\slashed\partial
  - e\gamma^\mu A^T_\mu - m)\psi_D \\
  &- \frac14 F_{\mu\nu}[A^T]F^{\mu\nu}[A^T]\Big].
\end{split}
  \label{eq:ST}
\end{equation}
Neither $S_{\rm T}$ nor the sources in Eq.~\eqref{eq:Z} contain $\xi$
or $\chi$. The integral over $\chi$ is Gaussian, factorizes and cancels
in normalized Green's functions. Every Green's function of $\psi_D$,
$\overline{\psi}_D$ and $A^T_\mu$ is therefore independent of $\xi$.

Its value can be read off at $\xi\to0$, where the Euclidean Gaussian
weight enforces $\partial^2\chi=0$ and hence $\chi=0$ for fields that vanish
at infinity. There $\psi_D=\psi$ and $A^T_\mu = A_\mu$, which proves
the Proposition,
\begin{equation}
  \langle \psi_D\cdots\overline{\psi}_D\cdots A^T\cdots\rangle_\xi
  = \langle \psi\cdots\overline \psi\cdots A\cdots\rangle_{\xi=0},
  \label{eq:proposition}
\end{equation}
for regularized Green's functions. In particular, the 1PI self-energy of
$\psi_D$ equals the Landau-gauge self-energy to all orders,
$\hat\Sigma^{(\psi_D)} = \Sigma_{\xi=0}$, of which
Eq.~\eqref{eq:result} is the one-loop case. Renormalization is
discussed in Sec.~\ref{sec:WI}.

The separation~\eqref{eq:split} is the all-orders form of the mechanism
of Sec.~\ref{sec:oneloop}. In momentum space, the tree-level coupling to
$A^T_\mu$ is the projected vertex~\eqref{eq:dressed_vertex}, and the
propagator of
$A^T_\mu$ is $D^T_{\mu\nu}$. The longitudinal part
$D^L_{\mu\nu} = -i\xi\,(k_\mu/k^2)(k_\nu/k^2)$ is the propagator of
$\partial_\mu\chi$, which never reaches the dressed sector.
Equation~\eqref{eq:DP_vanishes} is this statement for a single photon
line. The change of variables justifies the use of the projected vertex
at every vertex and to all orders, starting from local QED.

Three remarks are in order. First, the argument holds equally with a
photon-mass regulator, since
$A_\mu A^\mu = A^T_\mu A^{T\mu} + \partial_\mu\chi\,\partial^\mu\chi$
up to a total derivative, so a mass term also separates. Second, the
argument does not use the locality of Eq.~\eqref{eq:dressed_Lagrangian}.
The regularized theory is local QED, and its renormalization is that of
local QED in the Landau gauge. Third, there is no Gribov
ambiguity~\cite{Gribov:1977wm, Singer:1978dk} in QED, and the Landau-gauge limit is stable under renormalization, since
$\xi_B = Z_3\,\xi$.

\section{Ward identity, renormalization and nonlocality}
\label{sec:WI}

\subsection{Homogeneous Ward identity}

In the variables~\eqref{eq:change} the photon enters the
dressed-fermion sector only through $A^T_\mu = P^T_{\mu\nu}A^\nu$.
Every attachment of an external photon with momentum $k$ therefore
carries the projector $P^T(k)$, and by the Proposition the full 1PI
photon vertex $\Gamma^\mu_D$ of the dressed field is the transverse part
of the full Landau-gauge vertex $\Gamma^\mu_{\rm L}$,
\begin{equation}
\begin{split}
  \Gamma^\mu_D(p{+}k,p) &= P^{T\mu}{}_\nu(k)\,\Gamma^\nu_{\rm L}(p{+}k,p),\\
  k_\mu\Gamma^\mu_D(p{+}k,p) &= 0,
\end{split}
  \label{eq:homWI}
\end{equation}
to all orders. At tree level $\Gamma^\nu_{\rm L} = \gamma^\nu$, and
Eq.~\eqref{eq:homWI} reduces to the tree-level
vertex~\eqref{eq:dressed_vertex} and its transversality~\eqref{transv}.
The homogeneous Ward identity expresses the local gauge invariance of
$\psi_D$, whose gauge transformation acts only on $\chi$.

The dressed field still carries the global charge, since the conserved
current is unchanged, $\overline{\psi}_D\gamma^\mu\psi_D =
\overline \psi\gamma^\mu\psi$. Insertions of this current obey the usual
inhomogeneous Ward--Takahashi identity,
\begin{equation}
  k_\mu\Lambda^\mu(p{+}k,p) = S_{\rm L}^{-1}(p{+}k) - S_{\rm L}^{-1}(p),
  \label{eq:inhomWI}
\end{equation}
where $\Lambda^\mu$ is the amputated current vertex and
$S_{\rm L}^{-1}(p) = \slashed p - m_B - \Sigma_{\xi=0}(p)$ is the inverse
propagator of $\psi_D$, with $m_B$ the bare mass. Since the mass cancels
in the difference, the identity holds equally for the renormalized quantities below, where $Z_1 = Z_2$. The two identities
are compatible because the
projector in Eq.~\eqref{eq:homWI} removes the longitudinal part of the
vertex, which is the part fixed by Eq.~\eqref{eq:inhomWI}.

\subsection{Renormalization}

Landau-gauge QED is renormalizable, with $\psi^{\rm un} = (Z_2^{\rm L})^{1/2}\psi$, $A_{\mu}^{\rm un} = Z_3^{1/2}A_\mu$,
$m_B = Z_m m$ and $e_B = \mu^\epsilon Z_e e$ in $d=4-2\epsilon$
dimensions, where the Ward identity gives $Z_e Z_3^{1/2} = 1$. Since
$\chi$ is linear in the gauge field, $\chi^{\rm un} = Z_3^{1/2}\chi$ and
\begin{equation}
  e_B\chi^{\rm un} = \mu^\epsilon e\chi .
  \label{eq:echi}
\end{equation}
By the Proposition, the unrenormalized dressed field
$\psi_{D}^{\rm un} = e^{ie_B\chi^{\rm un}}\psi^{\rm un}$ is renormalized multiplicatively as
\begin{equation}
  \psi_{D}^{\rm un} = (Z_2^{\rm L})^{1/2}\,\psi_D ,
  \label{eq:ZD}
\end{equation}
in every gauge. Although $\psi_D$ contains arbitrarily many photon
fields and inverse d'Alembertians, it requires no new counterterm and
does not mix with other operators.

In a gauge $\xi\neq0$, where $\psi^{\rm un} = (Z_2^{(\xi)})^{1/2}\psi$,
Eq.~\eqref{eq:ZD} reads
\begin{equation}
  \psi_D = \left(\frac{Z_2^{(\xi)}}{Z_2^{\rm L}}\right)^{1/2}
  e^{i\mu^\epsilon e\chi}\,\psi .
  \label{eq:composite}
\end{equation}
The product $e^{i\mu^\epsilon e\chi}\psi$ of renormalized fields thus requires the
factor $(Z_2^{(\xi)}/Z_2^{\rm L})^{1/2}$, which is the gauge dependence
of the wave-function renormalization determined by the LKF
transformation~\cite{Johnson:Zumino:1959}. At one loop in minimal
subtraction $Z_2^{(\xi)} = 1 - \xi\alpha/(4\pi\epsilon)$, so
$Z_2^{\rm L}$ is finite at this order, consistent with
Eq.~\eqref{eq:result}. This is the only renormalization the dressing
requires.

\subsection{Nonlocality}

These results do not make $\psi_D$ local. The projector $P^T$, the
interaction in Eq.~\eqref{eq:dressed_Lagrangian} and the dressed field
itself are nonlocal in spacetime. What the Proposition shows is
that this nonlocality does not obstruct perturbative calculation or
renormalization, because the dressed Green's functions are those of local field theory in the Landau gauge.

The locality of the Landau-gauge field $\psi$ does not transfer to
$\psi_D$ as an operator. In a covariant gauge, $\psi$ is local but not
gauge invariant, and it acts in a space with indefinite metric. In a
positive-metric space where Gauss's law holds, physical charged states
cannot be created by local operators~\cite{Ferrari:1974rc, Buchholz:1986uj, Grassi:Porrati:2024}, and the nonlocality of a dressing
reflects this fact. Equal-time commutation relations, the asymptotic
behavior entering the Lehmann--Symanzik--Zimmermann reduction formula, causality and cluster
decomposition concern the operator $\psi_D$ itself, and we do not
address them here.

\section{Partially dressed fields and the LKF transformation}
\label{sec:LKF}

Consider the family
\begin{equation}
  \psi_\kappa \equiv e^{-i\kappa e\chi}\psi_D = e^{i(1-\kappa)e\chi}\psi
  \label{eq:psis}
\end{equation}
with real $\kappa$. Under a gauge transformation
$\psi_\kappa\to e^{-i\kappa e\Lambda}\psi_\kappa$, so $\psi_\kappa$ is
gauge invariant only at $\kappa=0$, and $\kappa=1$ is the undressed
field.

At tree level, $\psi_\kappa$ couples to the photon through the vertex
$\Gamma_{\kappa,0}^\mu = \gamma^\mu - (1-\kappa)\slashed k k^\mu/k^2$.
Each end of the
longitudinal propagator~\eqref{eq:photon_split} then contributes
$\Gamma_{\kappa,0}^\mu k_\mu/k^2 = \kappa\slashed k/k^2$, so the
longitudinal exchange is rescaled by $\kappa^2$, one factor $\kappa$ from
each vertex.
The partial dressing thus acts as a square root of a change of the
gauge parameter. The following factorization makes this exact. In the
variables of Sec.~\ref{sec:allorders}, $\chi$ is a free field decoupled
from
$\psi_D$, with propagator $\langle\chi\chi\rangle_\xi = \xi\,\Delta$,
\begin{equation}
  \Delta(x) = \int\frac{d^dk}{(2\pi)^d}\,
  \frac{-i\,e^{-ik\cdot x}}{(k^2+i\epsilon)^2},
  \label{eq:Delta}
\end{equation}
so that the two-point function factorizes,
\begin{equation}
\begin{split}
  \langle\psi_\kappa(x)\overline{\psi}_\kappa(y)\rangle_\xi
  &= \langle\psi_D(x)\overline{\psi}_D(y)\rangle \\
  &\quad\times e^{-\kappa^2e^2\xi\,[\Delta(0)-\Delta(x-y)]}.
\end{split}
  \label{eq:LKF}
\end{equation}
For $\kappa=1$, this is the LKF transformation from the Landau
gauge~\cite{Landau:Khalatnikov:1955,Fradkin:1955,Johnson:Zumino:1959,%
Zumino:1959wt}. Comparing the two, the propagator of $\psi_\kappa$ in the
gauge $\xi$ equals the propagator of $\psi$ in the gauge
\begin{equation}
  \xi' = \kappa^2\,\xi ,
  \label{eq:xiprime}
\end{equation}
exactly. The same holds for all Green's functions of $\psi_\kappa$, $\overline{\psi}_\kappa$ and $A^T_\mu$, since the Gaussian average of the
phases, which contains the propagator of $\chi$ between every pair of fermion insertions, depends on $\kappa$ and $\xi$ only through
$\kappa^2\xi$.\footnote{Equation~\eqref{eq:xiprime} is the relation $\alpha' = (1+X)^2\alpha$ obtained in Ref.~\cite{DeMeerleer:2018} from a
field-dependent gauge transformation, with $1+X = \kappa$.}
At one loop, where $\Sigma_\xi$ is linear in $\xi$,
the self-energy of $\psi_\kappa$ in the gauge $\xi$ is
\begin{equation}
\begin{split}
  \Sigma^{(\psi_\kappa)}_\xi(q) &= \Sigma_{\xi'}(q) \\
  &= \Sigma_{\xi=0}(q)
    + \frac{\xi'}{\xi}\bigl[\Sigma_\xi(q) - \Sigma_{\xi=0}(q)\bigr].
\end{split}
  \label{eq:SigmaP_s}
\end{equation}

A field redefinition and a change of the gauge parameter are different
operations. Equation~\eqref{eq:xiprime} is an identity between Green's
functions that relates the two, and it is exact in QED. The value $\kappa=0$
gives the unique gauge-invariant member of the family, where $\xi'=0$
for every $\xi$ and the Green's
functions no longer depend on the gauge. The map $\xi\to\xi'$ is
invertible for $\kappa\neq0$, while at $\kappa=0$ it sends every $\xi$ to
$\xi'=0$, so the dressed field carries no memory of the gauge in which
it is computed.

\section{Other dressings and Wilson lines}
\label{sec:other}

\subsection{General kernels}

A general linear dressing has the form
\begin{equation}
  \chi_V(x) = \int d^4y\, V^\mu(x-y)\,A_\mu(y),
  \label{eq:kernel}
\end{equation}
where $V^\mu$ is a nonlocal linear kernel rather than an ordinary
vector. Under $A_\mu\to A_\mu+\partial_\mu\Lambda$, with $\Lambda$
vanishing at infinity, the dressing shifts as $\chi_V\to\chi_V+\Lambda$,
so that $\psi_V = e^{ie\chi_V}\psi$ is gauge invariant, provided
\begin{equation}
  \frac{\partial}{\partial x^\mu} V^\mu(x-y) = \delta^4(x-y).
  \label{eq:kernel_condition}
\end{equation}
With the Fourier conventions
\begin{equation*}
\begin{split}
  A_\mu(x) &= \int\frac{d^4k}{(2\pi)^4}\,e^{-ik\cdot x}\tilde A_\mu(k), \\
  V^\mu(x) &= i\int\frac{d^4k}{(2\pi)^4}\,e^{-ik\cdot x}v^\mu(k),
\end{split}
\end{equation*}
the dressing becomes $\tilde\chi_V(k) = i v^\mu(k)\tilde A_\mu(k)$, and
the condition~\eqref{eq:kernel_condition} has the general solution
\begin{equation}
\begin{split}
  &k_\mu v^\mu(k) = 1, \qquad
  v^\mu(k) = \frac{k^\mu}{k^2} + v_T^\mu(k), \\
  &k_\mu v_T^\mu(k) = 0.
\end{split}
  \label{eq:kernel_solution}
\end{equation}
Gauge invariance thus fixes the longitudinal part of the kernel, while
the transverse part $v_T^\mu$ parametrizes the freedom in choosing
different dressings. The Dirac dressing~\eqref{chi} has
$V^\mu(x-y) = \partial G(x-y)/\partial x_\mu$, with $G$ the Green's function of
$\partial^2$ defined in Sec.~\ref{sec:transverse}, and $v_T^\mu = 0$.
Dirac's original kernel $\chi = \partial_iA_i/\nabla^2$ has
$v^\mu = [k^\mu - (u\cdot k)u^\mu]/[k^2 - (u\cdot k)^2]$, with
$u^\mu$ the unit vector in the time direction, and a straight Wilson
line joining $x$ to infinity along a fixed direction $n^\mu$ has
$v^\mu = n^\mu/(n\cdot k)$.

Since $\chi_V$ shifts by $\Lambda$, the Green's functions of $\psi_V$ can
be evaluated in the gauge $\chi_V = 0$, where $\psi_V=\psi$. They are
those of $\psi$ in that gauge, with the pole prescription fixed by
the kernel. The tree-level vertex of $\psi_V$ is
$\gamma^\mu - \slashed k\,v^\mu = \Gamma_{D,0}^\mu - \slashed k\,v_T^\mu$,
and the fermion couples to $(g_{\mu\nu}-k_\mu v_\nu)A^\nu$. Since
$(g_{\mu\nu}-k_\mu v_\nu)k^\nu = 0$ by Eq.~\eqref{eq:kernel_solution},
the longitudinal part of the photon propagator drops out for every
kernel. For kernels odd in $k$, $v^\mu(-k) = -v^\mu(k)$, as for the
Dirac and Coulomb kernels, the effective photon propagator is
\begin{equation}
\begin{split}
  &-\frac{i}{k^2}\left(g_{\mu\nu} - k_\mu v_\nu - v_\mu k_\nu
  + v^2\, k_\mu k_\nu\right) \\
  &\quad= D^T_{\mu\nu} + \frac{i}{k^2}\left(k_\mu v_{T\nu}
  + v_{T\mu}k_\nu - v_T^2\, k_\mu k_\nu\right),
\end{split}
  \label{eq:Vprop}
\end{equation}
the propagator of the gauge $v\cdot A = 0$. The transverse part of the
kernel enters only through terms proportional to $k_\mu$ or $k_\nu$.
For $v_T^\mu\neq0$ these terms remain in the off-shell self-energy, and
the dressing correction does not drop out as it does in
Sec.~\ref{sec:oneloop}. In this way Dirac's original kernel gives the
Coulomb gauge, and the Wilson line gives the axial gauge $n\cdot A=0$,
with the pole at $n\cdot k=0$ prescribed by the orientation of the line.

Different kernels therefore define physically different operators with
different off-shell Green's functions. They also differ in their infrared
behavior. Dressings adapted to the velocity of the charge admit an
infrared-finite mass-shell
renormalization~\cite{Bagan:Lavelle:McMullan:1996,Bagan:1996su}, in the
spirit of the asymptotic dressings of Kulish and
Faddeev~\cite{Kulish:Faddeev:1970}, while the Dirac
dressing~\eqref{chi} inherits the infrared structure of the Landau
gauge. What all dressings
share is the position of the mass-shell singularity, because the pole
mass is gauge independent~\cite{Nielsen:1975,Tarrach:1981,%
Breckenridge:Lavelle:Steele:1995,Kronfeld:1998,Gambino:Grassi:2000}.

\subsection{Wilson-line bilocals}

A different gauge-invariant object is the bilocal
\begin{equation}
  G_C(x,y) = \left\langle \overline \psi(y)\,
  e^{-ie\int_C dz^\mu A_\mu(z)}\,\psi(x)\right\rangle,
  \label{eq:WL_QED}
\end{equation}
with a contour $C$ from $x$ to $y$. It is the propagator of neither
$\psi$ nor $\psi_D$. Using $A_\mu = A^T_\mu + \partial_\mu\chi$ and
$\int_C dz^\mu\partial_\mu\chi = \chi(y)-\chi(x)$, one finds
\begin{equation}
  G_C(x,y) = \left\langle \overline{\psi}_D(y)\,
  e^{-ie\int_C dz^\mu A^T_\mu(z)}\,\psi_D(x)\right\rangle.
  \label{eq:WL_T}
\end{equation}
The bilocal thus differs from the dressed propagator by a
gauge-invariant transverse Wilson factor, which is not unity. Two
contours $C_1$ and $C_2$ differ by the closed loop
\begin{equation}
  e^{-ie\oint_{C_1-C_2} dz^\mu A_\mu}
  = e^{-\frac{ie}{2}\int_{\mathcal{S}} d\sigma^{\mu\nu}F_{\mu\nu}},
  \label{eq:loop}
\end{equation}
where $\mathcal{S}$ is a surface bounded by $C_1-C_2$. The contour dependence
is a physical flux, not a gauge artifact, and the results for the
self-energy of $\psi_D$ do not apply to $G_C$.

\section{The mass}
\label{sec:mass}

\subsection{Mass function of the dressed field}

The Landau-gauge self-energy~\eqref{eq:result} contains ultraviolet
divergences. The on-shell subtraction
\begin{equation}
\begin{split}
 m_D(q) = m &+ \hat\Sigma^{(\psi_D)}(q) - \hat\Sigma^{(\psi_D)}(m) \\
  &- (\slashed{q}{-}m)\frac{d\hat\Sigma^{(\psi_D)}}{d\slashed{q}}(m)
\end{split}
  \label{mass}
\end{equation}
removes them and defines the renormalized mass function of the dressed
field~\cite{Choi:2026yxc,Choi:2023cqs,Choi:2026edn}. The mass function
$m_D(q)$ is a combination of the unit matrix and $\slashed q$, and the
dressed propagator is $i/(\slashed{q}-m_D(q))$, with $m_D(m) = m$ and
unit residue. With a massless photon, the derivative in the last term is
infrared divergent in the Landau gauge, as is the conventional on-shell
wave-function renormalization. Off shell, $m_D(q)$ is therefore defined
at a fixed infrared regulator~\cite{Choi:2026edn}. The pole mass is
infrared finite.

\subsection{Comparison with the pinch technique}

For the undressed field, the pinch
technique~\cite{Cornwall:1982,Cornwall:Papavassiliou:1989,%
Papavassiliou:1995,Binosi:Papavassiliou:2009,%
Cornwall:Papavassiliou:Binosi:2011} rearranges the longitudinal parts
of the self-energy, vertex and box graphs of an amplitude and yields
the gauge-independent self-energy $\Sigma_{\xi=1}$. This holds at one
and two loops~\cite{Binosi:Papavassiliou:2002}. The pinch-technique
self-energy is a building block of amplitudes, not the propagator of
$\psi$. By Eq.~\eqref{eq:Omega} the dressed and pinch-technique
self-energies differ at one loop by
\begin{equation}
  \Sigma_{\xi=0}(q) - \Sigma_{\xi=1}(q) = (\slashed q - m)\,\Omega(q),
  \label{eq:two_masses}
\end{equation}
which vanishes at the pole, so both give the same pole mass. The pinch
technique extracts from the vertex and box graphs the longitudinal
contributions that the dressing places in the vertex from the start.
Writing $D_{\mu\nu} = D^T_{\mu\nu} + \xi D^{(1)}_{\mu\nu}$ with
$D^{(1)}_{\mu\nu} = -ik_\mu k_\nu/(k^2)^2$, the pinch technique removes
$(\xi-1)D^{(1)}_{\mu\nu}$ from the self-energy, while the dressing
removes $\xi D^{(1)}_{\mu\nu}$.

The difference~\eqref{eq:two_masses} is kinetic, and in any amplitude
it can be assigned to the vertices, where the Ward--Takahashi identity
ties it to a compensating shift of the vertex correction. With the
pinch-technique convention of Refs.~\cite{Choi:2026yxc,Choi:2026edn},
which assigns to the vertices every term the Ward--Takahashi identity
can act on, the dressing terms in $\Gamma_{D,0}^\mu$ and the $k_\mu k_\nu$
term of $D^T_{\mu\nu}$ are assigned there as well. At one loop every
dressing then leads to the same mass function, the one obtained from
Eq.~\eqref{mass} with $\Sigma_{\xi=1}$ in place of
$\hat\Sigma^{(\psi_D)}$. The propagator of $\psi_D$ itself keeps the
Landau form, and the two mass functions differ by
$(\slashed{q}-m)[\Omega(q)-\Omega(m)]$, carried by the vertex. Other assignments
select other reference gauges~\cite{Denner:Weiglein:Dittmaier:1994,%
Choi:2026edn}, while the pole and all amplitudes are the same for every
choice.

\section{Conclusions and outlook}

For the Dirac dressing~\eqref{chi} in QED, the Green's functions of the dressed
fermion and the transverse photon coincide, for any value of $\xi$ in
the $R_\xi$ gauge and to all orders, with the Landau-gauge Green's
functions of the undressed fields. The proof is a change of variables in local QED that confines
the gauge parameter to a decoupled sector. It implies a homogeneous
Ward identity for the dressed vertex, multiplicative renormalization of
the dressed field by the Landau-gauge constant and the exact LKF
relation~\eqref{eq:xiprime} for partially dressed fields. Dressings
with other kernels reproduce other gauges. They are physically
different operators that share the pole mass. Wilson-line bilocals are
distinct objects whose contour dependence is physical. The nonlocality
of the dressing remains, but it does not obstruct perturbative
calculation and renormalization. The infrared properties of the Dirac
dressing~\eqref{chi} are those of the Landau gauge, and an
infrared-finite description of charged asymptotic states requires
dressings adapted to the velocity of the
charge~\cite{Bagan:Lavelle:McMullan:1996,Bagan:1996su}.

We have restricted the analysis to QED. The Abelian construction does
not carry over directly to a non-Abelian theory. For
$U = e^{ig\chi^a t^a}$ with $\chi^a = \partial\cdot A^a/\partial^2$,
the transformed covariant derivative generates an infinite series of
commutator terms, and $\chi^a$ does not shift additively under gauge
transformations, so the transformed interaction is not the non-Abelian
analogue of Eq.~\eqref{eq:dressed_Lagrangian}. The appropriate
generalization is a dressing $h[A]$, constructed order by order, for
which the gauge field $A^h$ is transverse. The Green's functions of
$\psi^h = h^\dagger\psi$ are then independent of the gauge parameter
and equal to the Landau-gauge ones, and a local formulation with
$\psi^h$ is renormalizable to all
orders~\cite{DeMeerleer:2018,DeMeerleer:2020,Capri:2019drm}.
Nonperturbatively, the Gribov ambiguity~\cite{Gribov:1977wm,%
Singer:1978dk} obstructs a global dressing. An extension of the present
analysis along these lines is left for separate work.

\begin{acknowledgments}
The author thanks the anonymous referee for guidance. This work is
partly supported by the grant RS-2023-00277184 of the National Research
Foundation of Korea. The author used Claude (Anthropic) for editing and
discussion during the manuscript preparation. The final content was
reviewed and approved by the author, who takes full responsibility for
the work.
\end{acknowledgments}

\end{document}